\documentclass[aps,pre,showpacs,preprint]{revtex4}
\usepackage{graphicx,amsmath,bm}

\begin{document}
 
\title{Pair Approximation of the stochastic susceptible-infected-recovered-susceptible epidemic model 
on the hypercubic lattice}

\author{Jaewook Joo}
\affiliation{Department of Physics, Rutgers University, New Brunswick, New 
Jersey 08854, USA}
\author{Joel L. Lebowitz} 
\affiliation{Department of Mathematics and Physics, Rutgers University, New Brunswick, New 
Jersey 08854, USA}
\date{\today}

\begin{abstract}
We investigate the time-evolution and steady states of the stochastic 
susceptible-infected-recovered-susceptible~(SIRS) epidemic model
on one- and two- dimensional lattices. 
We compare the behavior of this system, obtained from computer simulations, 
with those obtained from the mean-field approximation~(MFA) 
and pair-approximation~(PA). The former~(latter) approximates higher order moments 
in terms of first~(second) order ones. 
We find that the PA gives consistently better results than the MFA. 
In one dimension the improvement is even qualitative.
\end{abstract}
\maketitle

\section{\label{sec:intro}Introduction}

The mathematical modeling of the spread of epidemics is a subject of continuing theoretical 
and practical interest~\cite{anderson:1992,diekmann:2000}. 
This is enhanced by the fact that the same or similar models are used for 
describing other phenomena such as plant and animal dispersal, and successional dynamics 
in ecology~\cite{neuhauser:2000,Jeger:1990}. 

The level of description provided by a model 
can be purely macroscopic and deterministic or individual and stochastic~\cite{durrett:1994a}. 
In the first case one uses (partial-) differential equations 
to describe the time evolution of different subpopulations; e.g., susceptible, 
infectious and recovered. In the second case one typically uses stochastic dynamics 
on a lattice (or more general graphs) where the variables at each node represent the state 
of an individual or a small spatial region. 
The time evolution of these variables is stochastic, e.g., an infected individual 
at site $i$ has a certain probability per unit time (rate) 
$\lambda$ to infect a susceptible individual at a neighboring site $j$. 
These systems fall into the category of what mathematicians call interacting particle 
systems~\cite{liggett:1985,durrett:1988b} and physicists call stochastic 
lattice gases~\cite{marro:1999} - systems of great interest also 
in the study of equilibrium phase transitions, phase segregation kinetics, etc., 
fields very different from epidemiology and ecology.

The connection between these modes of description and various intermediate ones 
has been investigated extensively in recent years, e.g., see ~\cite{lebowitz:1985-86a,
lebowitz:1986b,lebowitz:1988,durrett:1994,durrett:1994a,bramson:1997}. 
Mathematically this involves the use of the so called hydro-dynamical scaling limit. 
This uses a rigorous separation of space and time scales 
to derive deterministic macroscopic equations 
from the microscopic dynamics of stochastic lattice systems. 
Other approaches are based on more heuristic methods 
such as the mean field approximation (MFA) and improvement thereof~\cite{dickman:1986,
matsuda:1992,benavraham:1992,petermann:2004,levin:1995,levin:1996,levin:1997,durrett:1998,keeling:1999,
levin:2000,schnaiz:2002,geometry:2002}

The present work falls in the latter category. We apply a pair approximation (PA) 
scheme to a microscopic stochastic epidemic model in which individuals recovered 
from an infection 
enjoy a period of immunity before again becoming susceptible at a rate $\gamma$: 
the SIRS model. 
The PA approximation was used by Durrett and Levin~\cite{levin:1996} for the simpler 
susceptible-infected-susceptible~(SIS) model where 
recovered individuals immediately become susceptible again. 
They compared the results of the PA and MFA with those of the stochastic 
SIS model and found that the 
PA gave a quantitative improvement over the MFA. 
Here we consider the general SIRS model. 
We obtain the behavior of the stochastic model from extensive computer simulations. 
We then solve the PA and MFA models analytically for the stationary state and numerically 
for the time dependent case. We find that 
the PA gives considerably better agreement with the simulations than the MFA both for the 
time evolution and for the 
steady state. For the latter the PA reproduces the qualitative difference between 
the one and higher dimensional phase diagram of this model found in 
Ref.~\cite{kuulasmaa:1982, durrett:1991,andjel:1996,berg:1998}. This is reminiscent of 
the relation between the MFA and the Bethe-Peierls approximation
~(which the PA closely resembles) for equilibrium lattice systems~\cite{huang}.


\section{\label{sec:model}The stochastic SIRS model}

We first recall the stochastic lattice model of the SIRS epidemic process~\cite{murray:1980}.
A site $x$ of a $d$-dimensional lattice can be occupied by an individual in a state of  
$S$~(healthy and susceptible), $I$~(infected), or $R$~(recovered, i.e., healthy and immune).
The system evolves according to the following transition rates, 
\begin{eqnarray} 
S \rightarrow I &\text{ at rate }& \lambda n(x), \label{transition}
\\ \nonumber
I \rightarrow R &\text{ at rate }& \delta,
\\ \nonumber
R \rightarrow S &\text{ at rate }& \gamma, 
\end{eqnarray}
where $n(x)$ is the number of infected (nearest) neighbors of $x$, 
$\lambda$ is the infection rate, $\delta$ is the recovery rate 
and $\gamma$ is the rate at which immunization ceases.
The limit $\gamma \rightarrow \infty$ corresponds to the case where 
a recovered site passes instantaneously through the state $R$; 
this is the SIS model, also known as the contact process.
We shall choose time units in which $\delta$=1.

One can obtain some rigorous qualitative information about this and 
related models via probabilistic approaches such as those used in 
interacting particle systems~\cite{kuulasmaa:1982, durrett:1991, andjel:1996,berg:1998}. 
Of particular interest is the behavior of the 
stationary state on an infinite lattice which is a good approximation for 
the quasi-steady state behavior of large systems: see Appendix~\ref{MCsimulation}.
. This information is encoded in the phase 
diagram of the stationary state which depends on the infection rate $\lambda$, 
the recovery rate $\gamma$ and the topology of the lattice. For small $\lambda$, 
the only stationary 
state is one in which all sites are in the susceptible (disease-free) state
while for large $\lambda$ there is (for the infinite system) 
also a stationary state containing non zero fraction of $I$ and $R$ individuals.

The critical infection rate $\lambda_{c}(\gamma)$ is defined as the smallest value 
of $\lambda$, for a given $\gamma$, above which the infection can persist forever. 
For the SIS or contact process~($\gamma=\infty$), the critical infection value is 
known with high accuracy,
$\lambda_{c}(\infty) \simeq 1.6489$ in $d=1$ and 
$\lambda_{c}(\infty)\simeq 0.4122$ in $d=2$~\cite{liggett:1985,marro:1999}.
Considerably less is known about the phase diagram of the SIRS model.
Interestingly there is a qualitative difference in the behavior of 
$\lambda_{c}(\gamma)$ in one and in higher dimension when 
$\gamma \rightarrow 0$. 
It has been shown that $lim_{\gamma \rightarrow 0} \lambda_{c}(\gamma)=\lambda_{c}(0)$ 
is finite when $d \geq 2$ while $\lambda_{c}(0)=\infty$ when $d=1$
~\cite{kuulasmaa:1982, durrett:1991,andjel:1996,berg:1998}.

To go beyond qualitative results we need to carry out simulation or 
make some approximations. This is the subject of the rest of the paper.

\section{\label{sec:PA}The pair approximation}

The time evolution of the single site probabilities in the stochastic SIRS 
epidemic process can be written in the following form.
\begin{subequations}
\label{allmaster}
\begin{eqnarray}
\frac{dP_{t}(S_{x})}{dt}&=&-\lambda \sum_{y \in {\cal N}(x)} P_{t}(S_{x},I_{y})+\gamma 
P_{t}(R_{x}), 
\label{mastera}
\\ 
\frac{dP_{t}(I_{x})}{dt}&=&\lambda \sum_{y \in {\cal N}(x)} P_{t}(S_{x},I_{y})-P_{t}(I_{x}), 
\label{masterb}
\\ 
\frac{dP_{t}(R_{x})}{dt}&=&P_{t}(I_{x})-\gamma P_{t}(R_{x}),
\label{masterc}
\end{eqnarray} 
\end{subequations}
Here ${\cal N}(x)$ is the neighborhood~(nearest neighbor sites) of a site $x$, 
$P_{t}(\alpha_{x})$ is the probability of having a state 
$\alpha$ at site $x$ at time $t$ and $P_{t}(\alpha_{x},\beta_{y})$ is 
the joint probability to have state $\alpha$ at site $x$ and state 
$\beta$ at site $y$, at time $t$. We always have 
$P_{t}(S_{x})+P_{t}(I_{x})+P_{t}(R_{x})=1$.

Eqs.~(\ref{mastera})-~(\ref{masterc}) are, as is usual for moment equations, 
not a closed system. One can extend them by including 
equations for the time evolution of $P_{t}(S_{x},I_{y})$ which 
in turn involve higher moments of 
the spatial correlations. This leads to an infinite hierarchy. 
To solve such a hierarchy one 
usually resorts to some approximation scheme which expresses 
the higher order moments in terms of 
the lower order ones and truncates the equations at some point 
; this is referred to as the moment closure 
method~\cite{dickman:1986,matsuda:1992,benavraham:1992,petermann:2004,
levin:1995,levin:1996,levin:1997,levin:2000,
keeling:1999,geometry:2002,durrett:1998,schnaiz:2002}.
Both the MFA and PA are such schemes. In the MFA Eqs.~(\ref{mastera})-(\ref{masterc}) 
are closed by assuming that $P_{t}(S_{x},I_{y})=P_{t}(S_{x})P_{t}(I_{y})$, 
i.e., it neglects correlations between different sites. 
This leads to a pair of coupled equations which have been studied in \cite{murray:1980}.
In the PA scheme 
$P_{t}(\alpha_{x})$ and $P_{t}(\alpha_{x}, \beta_{y})$ are kept as unknowns while the 
higher-order moments are expressed, via some appropriate approximation, in terms of 
these quantities.

To carry out the PA we complement Eq.~(\ref{allmaster}) by equations 
for the second moments $P_{t}(\alpha_{x},\beta_{y})$
for nearest neighbor sites $x$ and $y$ based on the transition rule that we have 
described in Eq.~(\ref{transition}). These are 
\begin{subequations}
\label{alltwomoment}
\begin{eqnarray}
\frac{dP_{t}(S_{x},I_{y})}{dt} &=& \gamma P_{t}(R_{x},I_{y})- ( \lambda+1 
)P_{t}(S_{x},I_{y})+ 
\sum_{w \in {\cal N}^{x}(y)} \lambda P_{t}(S_{x},S_{y},I_{w}) 
\label{twomomenta}
\\ 
&-& \sum_{w \in {\cal N}^{y}(x)} \lambda P_{t}(I_{w},S_{x},I_{y}), 
\nonumber
\\ 
\frac{dP_{t}(S_{x},R_{y})}{dt} &=& P_{t}(S_{x},I_{y})+\gamma 
P_{t}(R_{x},R_{y})-\gamma 
P_{t}(S_{x},R_{y})-
\sum_{w \in {\cal N}^{y}(x)} \lambda P_{t}(I_{w},S_{x},R_{y}), 
\label{twomomentb}
\\ 
\frac{dP_{t}(R_{x},I_{y})}{dt} &=& 
-(\gamma+1)P_{t}(R_{x},I_{y})+P_{t}(I_{x},I_{y})+
\sum_{w \in {\cal N}^{x}(y)} \lambda P_{t}(R_{x},S_{y},I_{w})
\label{twomomentc}
\end{eqnarray}
\end{subequations}
where ${\cal N}^{x}(y)$ is the set of nearest neighbor sites of $y$ excluding the site $x$. 
$P_{t}(\alpha_{x},\beta_{y},\chi_{w})$ is the joint probability to have state $\alpha$ 
at site $x$, state $\beta$ at site $y$ and state $\chi$ at site $w$ at time $t$.
For a derivation of Eq.~(\ref{alltwomoment})
see Appendix~\ref{derivation_twomoment}.

To close the system~(\ref{allmaster}) and~(\ref{alltwomoment}) and derive a set of autonomous 
equations for $P_{t}(\alpha_{x})$ and 
$P_{t}(\alpha_{x},\beta_{y})$ we approximate the triad joint probability 
$P_{t}(\alpha_{x},\beta_{y},\chi_{w})$ for $x$ and $w$ nearest neighbors 
of $y$, by the product of two pair probabilities 
$P_{t}(\alpha_{x},\beta_{y})$ and $P_{t}(\beta_{y},\chi_{w})$ divided
by the probability $P_{t}(\beta_{y})$
~\cite{dickman:1986,matsuda:1992,levin:1996,benavraham:1992,petermann:2004}, i.e., we set
\begin{equation}
P_{t}(\alpha_{x}, \beta_{y}, \chi_{w}) = \frac{ P_{t}(\alpha_{x}, \beta_{y}) 
P_{t}(\beta_{y}, \chi_{w})} {P_{t}(\beta_{y})}
\label{myPA}
\end{equation}
Note that we have made use here of the structure of the hypercubic 
lattice. In such lattices, three adjacent sites, $x,y,w$ can not form a triangle 
but form only linear chains. This is not so in other lattices, 
e.g., the triangular lattice, where other configurations need also be considered.

While there are other choices for a PA, the approximation in Eq.~(\ref{myPA}) 
allows one to get the steady state solutions analytically. With other 
pair approximations~\cite{geometry:2002}, 
one has to solve the resulting differential equations numerically, making it 
impossible to obtain analytic expressions for the critical curve.

To actually carry out computations with the PA 
we will assume from now on that our system is spatially uniform.  
The site $x$, in Eqs.~(\ref{allmaster}) 
and~(\ref{alltwomoment}), can now be chosen to be the origin. 
We also define: $P_{t}(S,I)=\frac{1}{z}\sum_{y \in {\cal N}(x)} P_{t}(S_{x},I_{y})$, 
$P_{t}(\alpha,\beta,\chi)=\frac{1}{z-1}\sum_{w \in {\cal N}^{x}(y)}
P_{t}(\alpha_{x},\beta_{y},\chi_{w})$ where $z=2d$ is the number 
of nearest neighbors of a site in the $d$-dimensional cubic lattice. 
The truncated equations for the PA-SIRS can now be written, by using the exact 
Eqs.~(\ref{allmaster})-~(\ref{alltwomoment}) and the approximate Eq.~(\ref{myPA}), 
as a closed set of five coupled equations, 
\begin{subequations} 
\label{allPASIRS}
\begin{eqnarray}
\frac{d P_{t}(I)}{dt}&=& z \lambda P_{t}(S,I) -P_{t}(I) 
\label{PASIRSa}
\\
\frac{d P_{t}(R)}{dt}&=& P_{t}(I)-\gamma P_{t}(R) 
\label{PASIRSb}
\\
\frac{d P_{t}(S,R)}{dt}&=& P_{t}(S,I)+\gamma (P_{t}(R)-P_{t}(R,I)-2P_{t}(S,R))
-\frac{(z-1) \lambda P_{t}(S,I)P_{t}(S,R)}{1-P_{t}(R)-P_{t}(I)} 
\label{PASIRSc}
\\ 
\frac{d P_{t}(R,I)}{dt}&=& -(2+\gamma )P_{t}(R,I)+P_{t}(I)-P_{t}(S,I)
+\frac{(z-1) \lambda P_{t}(S,I) P_{t}(S,R)}{1-P_{t}(R)-P_{t}(I)} 
\label{PASIRSd}
\\
\frac{d P_{t}(S,I)}{dt}&=& \gamma P_{t}(R,I) -(\lambda+1) P_{t}(S,I)
\label{PASIRSe} 
\\
&+&\frac{(z-1) \lambda P_{t}(S,I)}
{1-P_{t}(I)-P_{t}(R)}(1-P_{t}(R)-P_{t}(I)-P_{t}(S,R)-2P_{t}(S,I)). \nonumber 
\end{eqnarray}
\end{subequations}
Note that we always have $P_{t}(\alpha)=P_{t}(\alpha,S)+P_{t}(\alpha,I)+P_{t}(\alpha,R)$ 
which determines 
$P_{t}(I,I)$ and $P_{t}(S,S)$.

In the limit $\gamma \rightarrow \infty$, 
$P_{t}(R)$ and $P_{t}(R,\alpha)$ as well as their time derivatives 
will go to zero. This yields $\gamma P_{t}(R)=P_{t}(I)$ and $\gamma P_{t}(R,I)=P_{t}(I)-P_{t}(S,I)$
~\cite{murray:1980}. 
In this limit Eq.~(\ref{allPASIRS}) reduces to the PA 
equations of the SIS~considered in~\cite{levin:1996},
\begin{subequations}
\label{PASIS}
\begin{eqnarray}
\frac{d P_{t}(I)}{dt}&=& z \lambda P_{t}(S,I) -P_{t}(I), 
\label{PASISa}
\\ 
\frac{d P_{t}(S,I)}{dt}&=& P_{t}(I)-(\lambda+2)P_{t}(S,I)
+\frac{(z-1)\lambda P_{t}(S,I)}{1-P_{t}(I)}(1-P_{t}(I)-2P_{t}(S,I)).
\label{PASISb}
\end{eqnarray}
\end{subequations}

As already noted the MFA approximates the joint probability 
$P_{t}(S,I)$ in Eq.~(\ref{PASIRSa}) by the product, $P_{t}(S,I)=P_{t}(S)P_{t}(I)$. 
This leads to the closed set of MFA of equations
for the SIRS~\cite{murray:1980},
\begin{subequations}
\label{MFPASIRS}
\begin{eqnarray}
\frac{dP_{t}(S)}{dt}&=&-z \lambda P_{t}(S)P_{t}(I)+\gamma 
P_{t}(R)
\label{MFPASIRSa} 
\\ 
\frac{dP_{t}(I)}{dt}&=&z \lambda P_{t}(S)P_{t}(I)-P_{t}(I)
\label{MFPASIRSb}
\\
\frac{dP_{t}(R)}{dt}&=&P_{t}(I)-\gamma P_{t}(R)
\label{MFPASIRSc}
\end{eqnarray}
\end{subequations}
For $\gamma \rightarrow \infty$, $\gamma P_{t}(R) \rightarrow P_{t}(I)$ and 
$P_{t}(S) \rightarrow 1-P_{t}(I)$. Eq.~(\ref{MFPASIRS}) then reduces to the MFA 
for the SIS.

\section{\label{stationary}Stationary solutions of the PA-SIRS model}

Let us first consider the steady state solutions of 
the PA-SIS obtained  by setting the l.h.s of Eq.~(\ref{PASIS}) 
equal to zero~\cite{levin:1996}. This gives for the critical value of the PA-SIS 
epidemic process $\lambda_{c}(\infty)=1/(z-1)$.
For $\lambda \leq \lambda_{c}(\infty)$, 
both $P_{t}(I)$ and $P_{t}(S,I)$ $\rightarrow 0$ as $t \rightarrow \infty$ 
for all initial states.
When $\lambda > \lambda_{c}(\infty)$ there is, in addition to the disease-free 
state corresponding 
to $P(I)=0$, also a stationary state consisting of a finite fraction of infected 
individuals:
\begin{subequations}
\label{stationarySIS}
\begin{eqnarray}
\bar{P}(S,I)&=&\bar{P}(I)/(z\lambda)
\label{StationarySISa}
\\ 
\bar{P}(I)&=&\frac{z[(z-1)\lambda-1]}{z(z-1)\lambda-1}.
\label{stationarySISb}  
\end{eqnarray}
\end{subequations}
It is these non-zero steady states which are approached as $t \rightarrow \infty$ 
when starting from any initial state with $P_{0}(I)>0$.

The steady state solutions of the PA-SIRS system 
is obtained by setting the l.h.s. of Eq.~(\ref{allPASIRS}) equal to zero. 
Setting $x=\bar{P}(I)$ this yields, 
\begin{subequations}
\label{steadyPASIRS}   
\begin{eqnarray}
\bar{P}(R) &=& x/\gamma  
\label{steadyPASIRSa}
\\
\bar{P}(S,I) &=& x/(z \lambda)
\label{steadyPASIRSb}
\\ 
\bar{P}(S,R) &=& \frac{x(\frac{1}{z
\lambda}+\frac{1}{\gamma+1})}
{\gamma(1+\frac{1}{\gamma+1}+\frac{(z-1)x}{z(\gamma-(1+\gamma)x)})}
\label{steadyPASIRSc}
\\
\bar{P}(R,I) &=& \frac{x-\gamma 
\bar{P}(S,R)}{\gamma+1} \\ \nonumber 
               &=& \frac{x}{\gamma+1}
\Bigl (
1-\frac
{
\frac{1}{z \lambda}+\frac{1}{\gamma+1}
}
{
1+\frac{1}{\gamma+1}+\frac{(z-1)x}{z(\gamma-(\gamma+1)x)}
}
\Bigr )
\label{steadyPASIRSd}
\end{eqnarray} 
\end{subequations}
where $\bar{P}(\alpha,\beta)$ are the approximate probabilities for having states 
$\alpha$ and $\beta$ on neighboring sites. After further simplifications, 
we find that $x$ has to satisfy the cubic equation,
\begin{equation}
x(a_{1} x^{2}+a_{2} x+a_{3} )=0
\label{Pequation}
\end{equation}
Both the derivation of Eq.~(\ref{Pequation}) and the explicit  expressions for 
$a_{1}$, $a_{2}$ and $a_{3}$ as functions of $\lambda$ and $\gamma$ are given in 
Appendix~\ref{Pcoefficients}.

The root $x=0$ corresponds to the all healthy steady state, which is always 
a solution. The critical curve $\lambda_{c}(\gamma)$ is determined by 
the existence of a root of Eq.~(\ref{Pequation}) such that $x$ and all other 
stationary probabilities, are strictly positive.
It turns out that this strictly positive root is unique.
Thus when $\lambda \leq \lambda_{c}(\gamma)$, 
$x=0$ is the only steady state solution. For $\lambda> \lambda_{c}(\gamma)$, 
there is also a steady state in which the infection is endemic:
$\bar{P}(I)=\gamma \bar{P}(R)=x$ and $\bar{P}(S)=1-(1+1/\gamma)x$, see Appendix
~\ref{Pcoefficients}.

The critical curve $\lambda_{c}(\gamma)$ is obtained in Appendix~\ref{Pcoefficients}. 
It is given by the equation,  
\begin{equation}
\lambda_{c}(\gamma)=\frac{\gamma+1}{2d-2+(2d-1)\gamma}, \textrm{  d=1,2,3,...}
\end{equation}
As $\gamma \rightarrow \infty$, $\lambda_{c}(\infty)=(2d-1)^{-1}$, the critical 
point of the PA-SIS epidemic 
process. On the other hand, as $\gamma$ approaches zero,
the critical curve shows different behavior depending on the 
dimension of the lattice: $\lambda_{c}(0)$ diverges to infinity for $d=1$, 
while $\lambda_{c}(0)$ is finite for $d\geq2$.
The PA thus reproduces the qualitative difference between the one and higher 
dimensional phase diagram of the SIRS model found in Ref.
~\cite{kuulasmaa:1982, durrett:1991,andjel:1996,berg:1998}.

The MFA, Eq.~(\ref{MFPASIRS}), yields the mean field critical value, $\lambda^{MF}_{c}=1/z$ 
independent of $\gamma$. 
In the coexistence region $\lambda>\lambda^{MF}_{c}$ the mean field stationary states are  
$\bar{P}(I)=\gamma \bar{P}(R)=\frac{\gamma (\lambda z-1)}{\lambda z(\gamma+1)}$ and 
$\bar{P}(S)=\frac{1}{z \lambda}$.

\begin{figure}
\begin{center}
\includegraphics[height=10cm,width=10cm,angle=0]{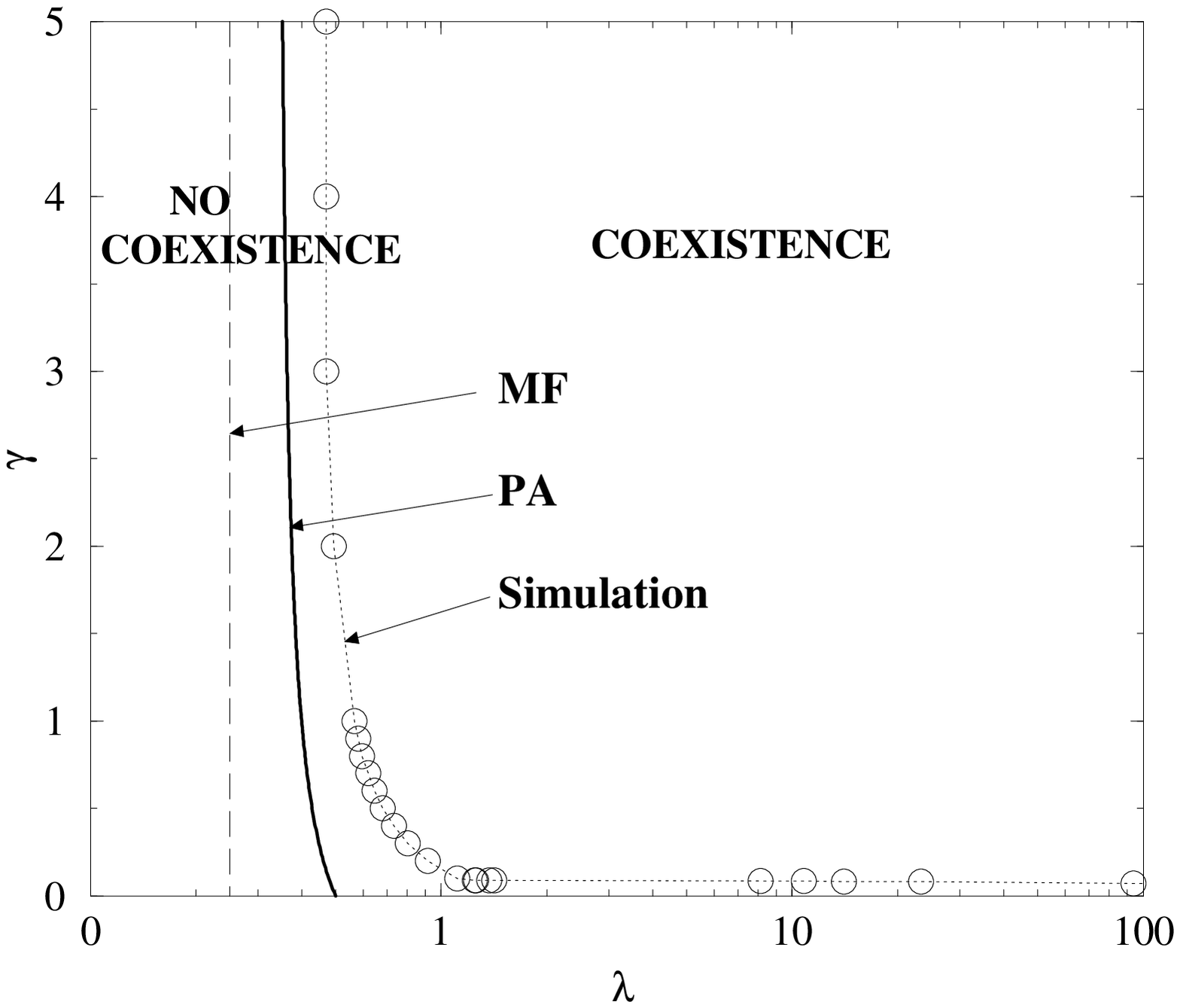}
\caption{\label{fig1}Phase diagram of the SIRS process in two dimensions.
The coexistence phase of $S$-$I$-$R$ and the no-coexistence phase are 
separated by the critical curve from the simulation~(open circles with dotted line for 
eye-guidance), the PA (thick solid line) and the MFA (long dashed line). 
The critical curve is obtained on periodic square lattice of different sizes $N$ 
from simulations extrapolated to infinite system : $N=50^{2},70^{2},100^{2},150^{2},200^{2}$. 
}
\end{center}
\end{figure}

\begin{figure}
\begin{center}
\includegraphics[height=10cm,width=10cm,angle=0]{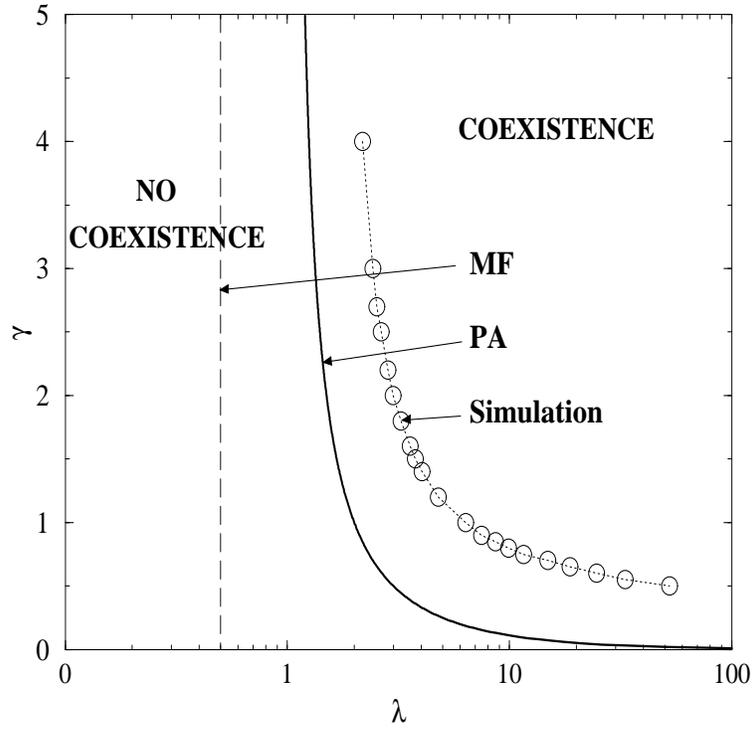}
\caption{\label{fig2}Phase diagram of the SIRS process in one dimension. 
The critical curve from numerical simulations of ring lattice of different sizes $N$ 
is extrapolated to infinite system: $N=5000,7000,10000,15000$. The same symbols are used as 
in the Fig.~\ref{fig1}. 
}
\end{center}
\end{figure}

Both the steady state and critical value of the MFA and PA fail to correctly represent 
the results of the stochastic SIRS process for small $\gamma$: 
see Figs.~\ref{fig1} and \ref{fig2}. 
Note in particular that $\bar{P}(S)$ of the stochastic SIRS process is
considerably larger than that of the MFA or PA for large $\lambda$ and small $\gamma$. 
This is due to the fact that the susceptible sites can be surrounded by recovered ones 
and thus protected from contacting infected ones in the stochastic case.

\section{Comparison of the stochastic, the  PA and MFA steady states.}

\begin{figure}
\begin{center}
\includegraphics[height=10cm,width=10cm,angle=0]{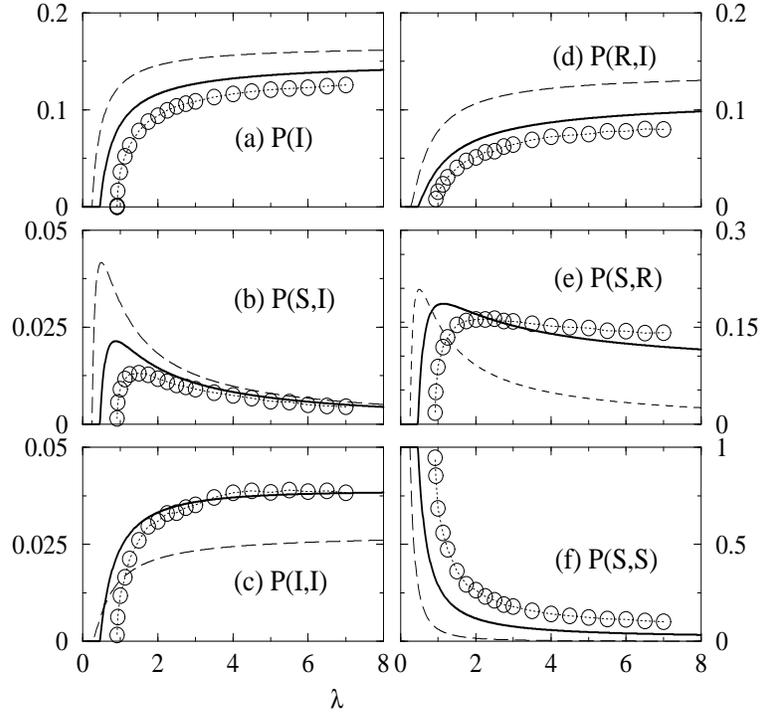}
\caption{\label{fig3}First and second order moments of the steady state SIRS 
in two dimensions at $\gamma=0.2$. 
The steady-state values of the density of infection in Fig.~\ref{fig3}(a) 
and the second moments in Fig.~\ref{fig3}(b)-(f) 
are drawn from the numerical simulation~(open circle with dotted line 
for eye-guidance), the PA~(thick solid line), and the MFA~(long-dashed line). 
For the numerical simulation we used a system of size $N=100^{2}$.}
\end{center}
\end{figure}

\begin{figure}
\begin{center}
\includegraphics[height=10cm,width=10cm,angle=0]{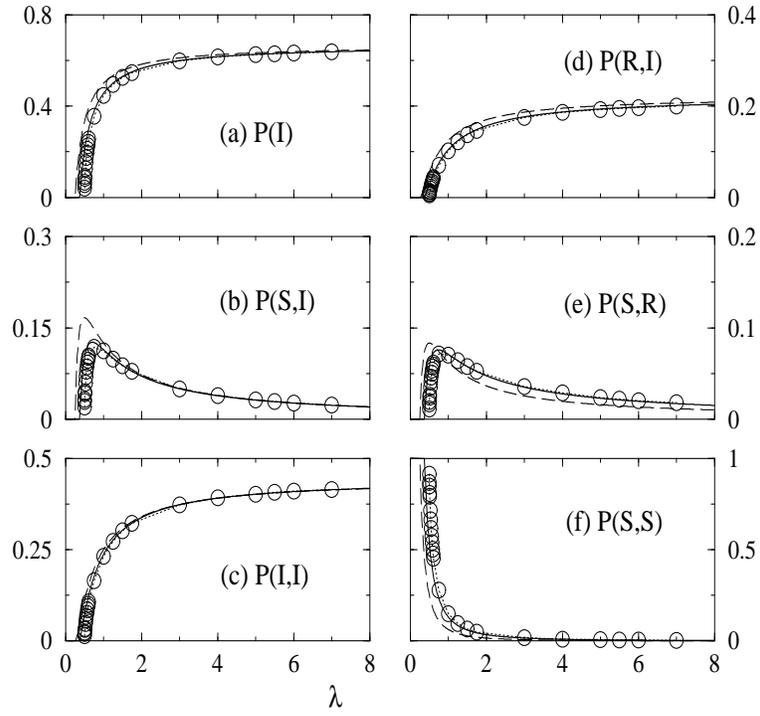}
\caption{\label{fig4}First and second order moments of the steady state SIRS 
in two dimension at $\gamma=2$. The same symbols are 
used as in the Fig.~\ref{fig3}.}
\end{center}
\end{figure}

\begin{figure}
\begin{center}
\includegraphics[height=10cm,width=10cm,angle=0]{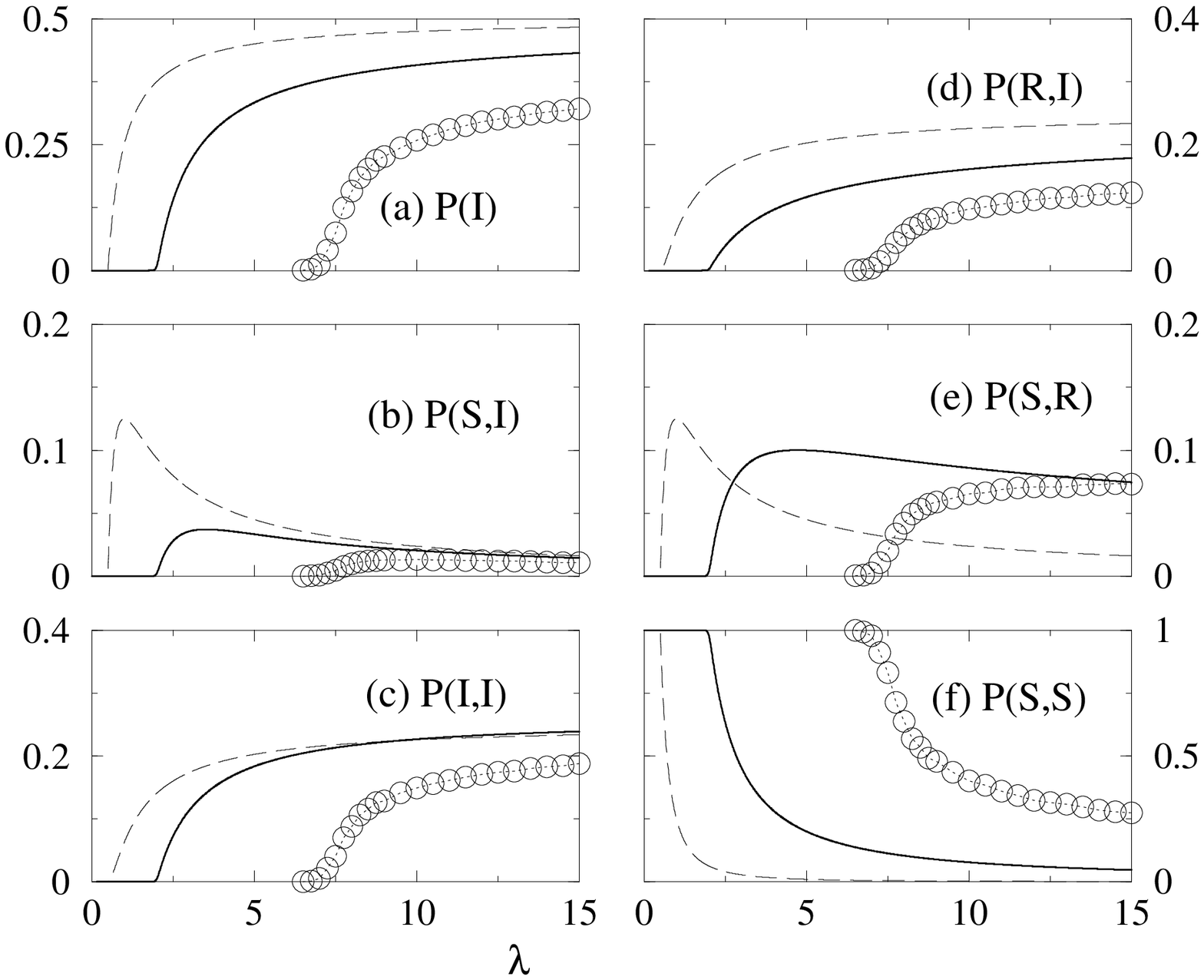}
\caption{\label{fig5}First and second order moments of the steady state SIRS 
process in one dimension at $\gamma=1$. 
The same symbols are used as in the Fig.~\ref{fig3}.}
\end{center}
\end{figure}

\begin{figure}
\begin{center}
\includegraphics[height=10cm,width=10cm,angle=0]{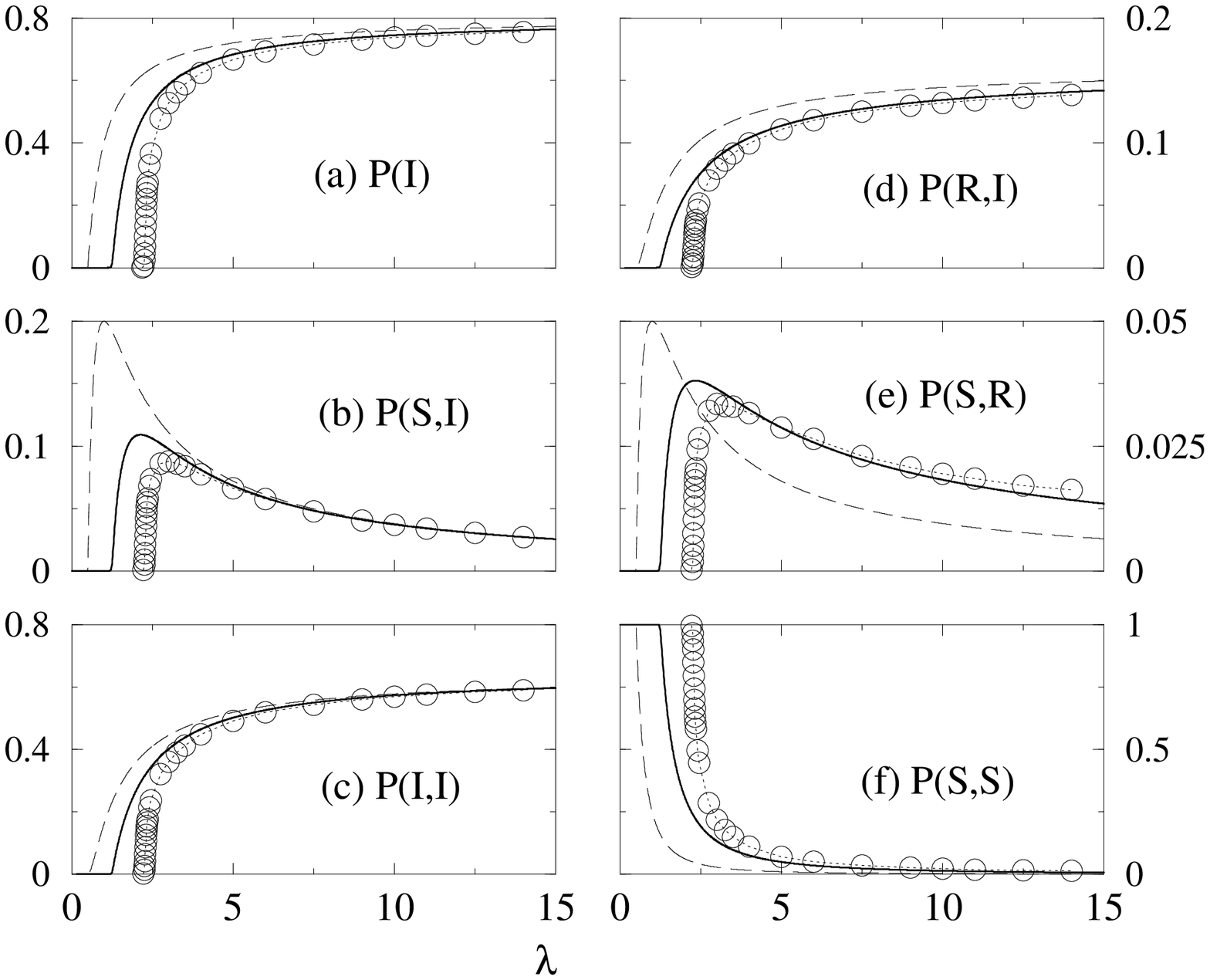}
\caption{\label{fig6}First and second order moments of the steady state SIRS 
process in one dimension at $\gamma=4$. The same 
symbols are used as in the Fig.~\ref{fig3}.}
\end{center}
\end{figure}

We compare in Figs.~\ref{fig3}~-~\ref{fig6} the steady state values of 
$\bar{P}(\alpha)$ and $\bar{P}(\alpha,\beta)$ obtained from the MFA and PA 
with the results from the stochastic SIRS process
as a function of $\lambda$ at fixed values of $\gamma$.
Clearly the PA gives results closer to those obtained from the stochastic model.
For the methods used to obtain the steady state results from the numerical simulation, 
see appendix.~\ref{MCsimulation}.

Figs.~\ref{fig3} and~\ref{fig5} show that both the MFA and PA overestimate  
$\bar{P}(I)$ as well as $\bar{P}(\alpha,I)$, $\alpha=S,R$.
This is due to the strong tendency of infected sites in the stochastic model 
to cluster into localized islands, reducing the contacts between $S$ and $I$.
This is partially taken into account by the PA as seen by the behavior of $\bar{P}(S,I)$ and 
$\bar{P}(I,I)$ 
in Figs.~\ref{fig3} and~\ref{fig5}. This clustering effect is also observed in the stochastic 
SIS process~\cite{levin:1996}. It is more pronounced in one dimension.

Note that $\bar{P}(S,I)$ becomes zero both at 
$\lambda<\lambda_{c}(\gamma)$, when $\bar{P}(I)=0$, and 
at $\lambda=\infty$ when $\bar{P}(S)=0$, reaching 
a peak at a positive value of $\lambda$ which depends on $\gamma$. 
For large values of $\gamma$, the steady state values of $\bar{P}(\alpha)$ and 
$\bar{P}(\alpha,\beta)$ obtained from the PA, or the MFA agree well with the 
numerical simulation, away from the critical $\lambda_{c}(\gamma)$. 
Moreover the PA yields steady state curves remarkably similar to those from 
the numerical simulation, see Figs.~\ref{fig4} and~\ref{fig6}.

\section{Linear stability analysis of the Pair approximation}

To study the stability of the stationary PA state, 
Eq.~(\ref{allPASIRS}) is linearized about the steady state values
~\cite{murray:1980}, see Appendix~\ref{PAjacobian}. 
This leads to the study of the roots of the characteristic fifth order polynomial $P_{5}(\xi)$, 
obtained from $|A-\xi I |=0$ where $A$ is the Jacobian of the linearized PA-SIRS system.
If $Re \xi <0$, the solution of the linearized equation is stable, i.e., 
a small perturbation around the steady state will decay back 
to the steady state. 
We used the Routh-Hurwitz conditions~\cite{murray:1980} to obtain the sign of 
the real part of eigenvalues of the Jacobian.  
As expected, the positive steady state solution is stable for $\lambda>\lambda_{c}(\gamma)$. 
The zero steady state solution is stable for $\lambda \leq \lambda_{c}(\gamma)$ 
and unstable for $\lambda > \lambda_{c}(\gamma)$.

The eigenvalues of $P_{5}(\xi)$ have non-zero 
imaginary parts in some regions of the parameter space. 
In such regions the PA-SIRS system in 
Eq.~(\ref{allPASIRS}) will converge to the steady state in a damped 
oscillatory manner. Such oscillations are seen in 
Fig.~\ref{fig7} and \ref{fig8}.

\section{Time dependent behavior}

\begin{figure}
\begin{center}
\includegraphics[height=10cm,width=10cm,angle=0]{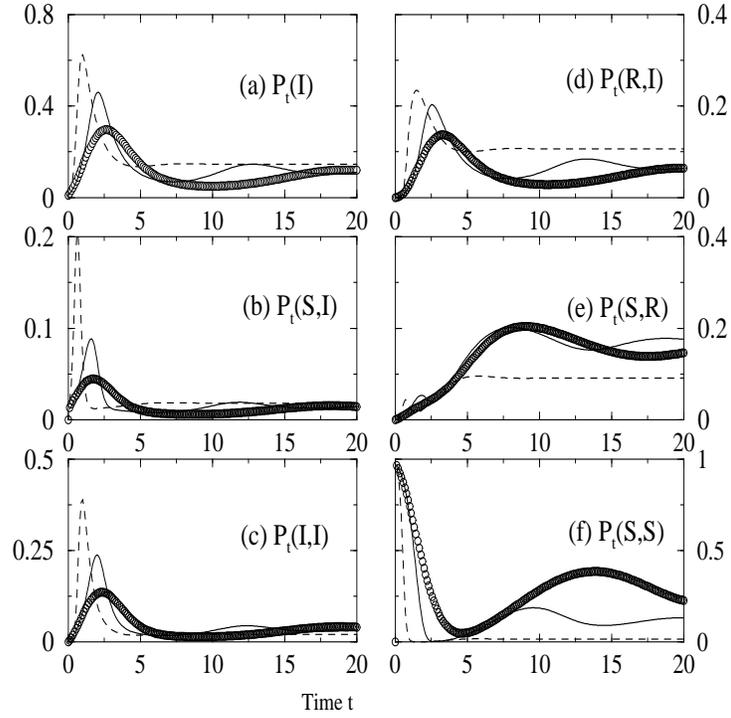}
\caption{\label{fig7}Time-evolution of the first and the second order moments of 
the SIRS process in two dimensions. All sub-graphs are from numerical 
simulations (open circles), the PA~(solid line), and the MFA~(dashed line) 
at $\gamma=0.2$ and $\lambda=2$. A periodic square lattice of $N=10^{4}$ sites is used 
in the numerical simulations averaged over $10^{3}-10^{4}$ realizations starting with 
random initial distribution with 1 percent of infected sites. }
\end{center}
\end{figure}

\begin{figure}
\begin{center}
\includegraphics[height=10cm,width=10cm,angle=0]{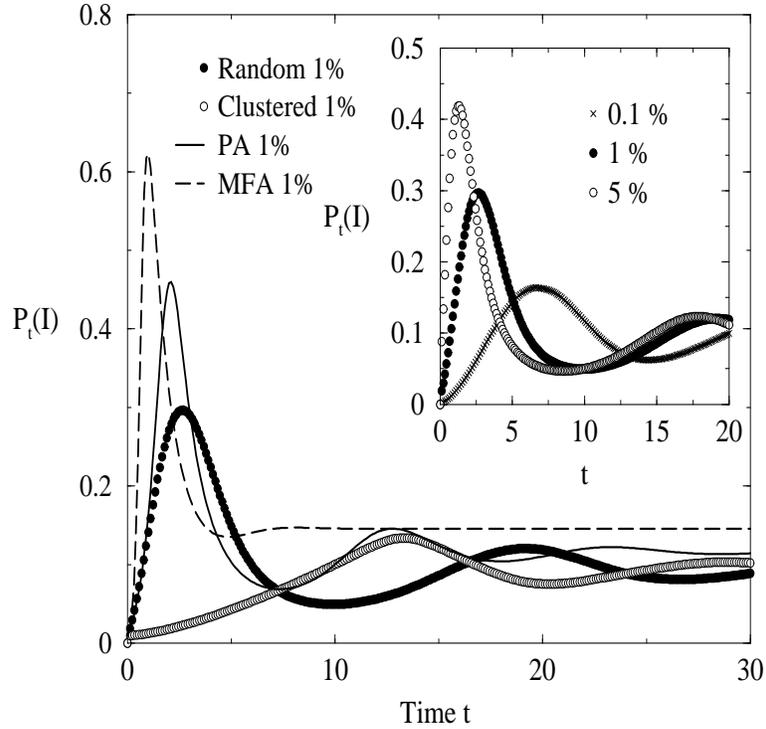}
\caption{\label{fig8}Time-evolution of fraction of infected sites of the 
SIRS process in two dimensions at $\gamma=0.2$ and $\lambda=2$. 
A periodic square lattice of $N=100^{2}$ is used in numerical 
simulation averaged over $10^{3}-10^{4}$ realizations.
Main: Simulation starts with 1$\%$ of infected sites placed either randomly
~(filled circles) or in a single cluster~(open circles) on a lattice.
Both the PA and MFA takes an initial value 0.01 for 
$P_{0}(I)$. 
Inset: Simulation starts with different fractions of infected sites 
randomly placed in a lattice: 0.1, 1 and 5 percents of the system.}
\end{center}
\end{figure}

To study the time evolution of an epidemic following an initial infection of 
a healthy population we performed dynamical 
Monte Carlo simulations~\cite{grassberger:1979} as well as solutions of 
Eqs.~(\ref{allPASIRS}) and (\ref{MFPASIRS}). For the stochastic evolution we started 
with infected sites placed either randomly or in a cluster and followed 
the time evolution averaged over $10^{3}$ realizations of the SIRS process.
To obtain the time evolution of the MFA and PA we solved Eq.~(\ref{MFPASIRS}) 
and Eq.~(\ref{allPASIRS}) numerically by using a 4th-order 
Runge-Kutta method. We plot the results in Figs.~\ref{fig7} and~\ref{fig8}.

To set the unit of time of the simulation 
we started with a fully infected state, $P_{0}(I)=1$ and $\lambda=0$ and obtained 
the exponentially decaying pattern of $P_{t}(I)$. We then set the slope (death rate) 
of the graph, 
$log P_{t}(I)$ vs $t$, from the numerical simulation equal to those from the MFA and PA.

Starting with a small value of $P_{0}(I)$, $P_{t}(I)$ displays  
an initial "exponential" growth in both the MFA and PA. 
Similar growth patterns are observed in all $P_{t}(\alpha,I)$, $\alpha=S,I,R$.
This is explained by the initially abundantly available susceptible population. 
Once the susceptible 
population is reduced, the infected population reaches a maximum and then 
decreases to the steady state 
endemic level. Note the damped oscillatory pattern in Figs.~\ref{fig7} and \ref{fig8} 
for this choice of the parameters ($\lambda$,$\gamma$).

The numerical simulation of the stochastic time evolution does not show the pronounced 
growth patterns of the PA and MFA when the initial fraction of infected sites is small, 
as seen in Fig.~\ref{fig8}. The formation of clusters of infected sites makes the 
infected population grow more slowly in the stochastic model. 
When the initial fraction of infected population 
increases to more than one percent the stochastic model shows significant change 
in its growth pattern, becoming similar to the PA and MFA. 
If however the same fraction of infected sites are initially placed in a single cluster 
the stochastic epidemic process exhibits slower growth patterns, similar to those starting with 
a small fraction of initially infected sites. These studies confirm that the clustering 
of infected sites in the stochastic model reduces both the speed of growth 
and the maximum fraction of infected sites. In realistic situations the population 
is not well mixed so we would expect 
growth patterns more similar to that of the stochastic epidemic model, starting with 
a fraction of infected sites initially placed in a single cluster.

\section{Summary}

We investigated the stochastic SIRS epidemic process and compared the results with those
obtained from the deterministic MFA and PA. These approximations close the hierarchy of  
dynamical equations by expressing the higher order moments in terms of the lower order ones.
The PA is found to improve over the MFA both for the stationary and 
for the time dependent states. The time evolution 
of the system shows damped oscillatory 
behavior in some parameter ranges.

\begin{acknowledgments}
Work supported by NSF DMR-01-279-26 and by AFOSR AF 49620-01-1-0154 and  
by DIMACS grants NSF DBI 99-82983 and NSF EIA 02-05116.
\end{acknowledgments}

\appendix

\section{\label{derivation_twomoment}Derivation of differential 
equation for $P_{t}(S_{x},I_{y})$}

Eq.~(\ref{twomomenta}) is derived by considering all transitions leaving or entering the pair 
configuration $(S_{x},I_{y})$. We list them as follows: A pair $(R_{x},I_{y})$ 
changes to a pair $(S_{x},I_{y})$ with a rate $\gamma$. A pair $(S_{x},I_{y})$ changes to 
a pair $(I_{x},I_{y})$ with a rate $\lambda$ and also changes to a pair $(S_{x},R_{y})$ 
with a rate 1. 
A triad configuration $(S_{x},S_{y},I_{w})$ 
transits to a triad $(S_{x},I_{y},I_{w})$ with 
a rate $\lambda$ such that a pair configuration $(S_{x},S_{y})$ is changed to $(S_{x},I_{y})$. 
A triad $(I_{w},S_{x},I_{y})$ 
changes to a triad $(I_{w},I_{x},I_{y})$ with a rate $\lambda$. 
The equations for $P_{t}(S_{x},R_{y})$ and $P_{t}(R_{x},I_{y})$ 
in Eq.~(\ref{alltwomoment}) can be obtained in a similar way.
The relation $P_{t}(\alpha_{x})=P_{t}(\alpha_{x},\alpha_{y})
+P_{t}(\alpha_{x},\beta_{y})+P_{t}(\alpha_{x},\chi_{y})$ can be used to obtain the 
other joint probabilities $P_{t}(\alpha_{x}, \beta_{y})$ 
which are not shown in Eq.~(\ref{alltwomoment}).

\section{\label{Pcoefficients}Derivation of Eq.~(\ref{Pequation})}
The steady states in Eq.~(\ref{steadyPASIRS}) are obtained by setting the l.h.s. of 
Eq.~(\ref{PASIRSa})-(\ref{PASIRSd}) equal to zero. In addition we set Eq.~(\ref{PASIRSe}) 
equal to zero
 and replace a single site and joint probabilities with the steady states in 
Eq.~(\ref{steadyPASIRS}). After simplifications, we obtain Eq.~(\ref{Pequation})
with the coefficients, 
\begin{eqnarray} 
a_{1}&=&
\gamma^{3} \{ z^{2}(z-1)\lambda-z \}
+\gamma^{2} \{ z(2z^{2}-2z-1)\lambda -2z-1\} \label{Pcoefficient}
\\ \nonumber 
&+&\gamma \{ 2z(z^{2}-z-1) \lambda -2z-1 \} 
+z \{(z^{2}-z-1)\lambda-1\}\\ \nonumber 
a_{2}&=&
z\gamma \Bigl \{
\gamma^{2} \{ z+1-2z(z-1)\lambda \}
+\gamma \{z+3-(3z^{2}-4z-1)\lambda \}
+z+1-(2z^{2}-3z-1)\lambda
\Bigr \}\\ \nonumber
a_{3}&=&
z^{2}\gamma^{2} \Bigl \{
\gamma\{-1+\lambda(z-1)\}-1+\lambda(z-2)
\Bigl \}.
\end{eqnarray}

The critical curve $\lambda_{c}(\gamma)$ is given by setting 
$a_{3}=0$. Only for $\lambda>\lambda_{c}(\gamma)$ does the quadratic factor of 
Eq.~(\ref{Pequation}) have a positive root. 

\section{\label{PAjacobian} The Jacobian of the linearized PA-SIRS}
The Jacobian of the linearized PA-SIRS is written,
\[
\mathbf{A}=
\left(
\begin{array}{ccccc}
-\gamma & 1 & 0 & 0 & 0 \\
0 & -1 & z\lambda & 0 & 0 \\
-K_{2}K_{0} 
& -K_{2}K_{0}
& K_{3}
& -\frac{K_{1}}{\bar{P}(SR)} & \gamma \\
\gamma-K_{2} & -K_{2} & 1-\frac{K_{1}}{\bar{P}(IS)} 
& -2\gamma-\frac{K_{1}}{\bar{P}(SR)} & -\gamma \\
K_{2} &  1+K_{2}  & -1+\frac{K_{1}}{\bar{P}(IS)} 
&  \frac{K_{1}}{\bar{P}(SR)}  &  -\gamma-2
\end{array} \right)
\]
where 
$K_{0}=1+2\frac{\bar{P}(IS)}{\bar{P}(SR)}$, 
$K_{1}=\frac{(z-1)\lambda 
\bar{P}(IS)\bar{P}(SR)}{1-\bar{P}(R)-\bar{P}(I)}$, 
$K_{2}=\frac{(z-1)\lambda
\bar{P}(IS)\bar{P}(SR)}{(1-\bar{P}(R)-\bar{P}(I))^{2}}$, and  
$K_{3}=(z-2)\lambda-1-K_{1}(\frac{1}{\bar{P}(IS)}+\frac{4}{\bar{P}(SR)})
$.

\section{\label{sec:stabilityMFSIRS}Linear Stability Analysis of the MF-SIRS}
The Jacobian matrix B of linearized MF-SIRS is given by~\cite{murray:1980}
\[
\mathbf{B}=
\left( \begin{array}{cc}
-\lambda z \bar{P}(I)-\gamma & -\lambda z \bar{P}(S)-\gamma \\
\lambda z \bar{P}(I) & \lambda z \bar{P}(S)-1 
\end{array}
\right)
\]
The characteristic polynomial of the second order, $P_{2}(\xi)=\xi^{2}+a_{1}\xi+a_{2}=0$, 
is obtained from $|B-\xi I|=0$.

The necessary and sufficient (Routh-Hurwitz) conditions~\cite{murray:1980} 
for $Re\xi<0$ is $a_{2}>0$ and $a_{1}>0$.
In the coexistence region where $z\lambda> 1$, 
$a_{2}=\gamma (z \lambda -1)>0$ and 
$a_{1}=\frac{\gamma}{\gamma+1}(\gamma+z\lambda)
>0$ for all $\gamma>0$.
In the no-coexistence region where 
$z\lambda<1$, 
$a_{2}=\gamma(1-z\lambda)>0$ and $a_{1}=\gamma +(1-z \lambda)>0$ 
for all $\gamma>0$. Both in the coexistence and 
no-coexistence region, the real part of the eigenvalues is negative 
and thus the mean field steady states are stable.

Now we turn our attention to the oscillatory behavior.
The eigenvalues of the characteristic polynomial $P_{2}(\xi)$ is given by, 
\begin{equation}
\xi_{\pm}=\frac{-\gamma^{2}-z \gamma \lambda \pm 
\sqrt{ (2\gamma z^{2}\lambda^{2}-2z\lambda (\gamma^{2}+2z\gamma \lambda 
+2)+\gamma^{3}+4\gamma^{2}+8\gamma+4) \gamma }}
{2(\gamma+1)}
\end{equation}
In the range of $\lambda_{-}(\gamma)<\lambda(\gamma)<\lambda_{+}(\gamma)$
the imaginary part of the eigenvalues is non-zero: 
$\lambda_{\pm}(\gamma)=\frac{2+4\gamma+\gamma^{2}\pm 
2(1+\gamma)^{3/2}}{z\gamma}$. In this range of $\lambda$, the steady states 
correspond to the stable spiral and the system converges to the steady state in 
damped oscillatory pattern. Even in the damped oscillatory region, any 
oscillation is hardly visible in the large $\gamma$ limit and becomes noticeable 
only in small $\gamma$ limit.

\section{\label{MCsimulation}Monte Carlo simulation}

The numerical simulations described here used lattices with 
periodic boundary conditions. In one dimension, rings of 
$5000 \leq N \leq 15000$ sites were used. 
In two dimensions, torii of $50^{2} \leq N \leq 200^{2}$ sites were employed.

To obtain the steady state of the SIRS process a random initial 
configuration of susceptible and infected sites is evolved according to 
the transition rates in Eq.~(\ref{transition}). 
In practice a site is randomly chosen and a random number~($\in [0,1]$) is 
also chosen: if it is greater than the given transition probability 
for that site, which is equal 
to the rate$\times \Delta t$, its state is updated: 
$\Delta t$ is chosen to be so small that transition
probability is not greater than 1 for a range of ($\lambda, \gamma$)
~\cite{marro:1999, durrett:1994b}. Otherwise its state remains the same.

For a finite system the only true stationary state of the SIRS process is 
the absorbing state corresponding to $P(S)=1$, $P(I)=P(R)=0$. To learn about 
the active state from simulations of 
a finite system we study the quasi-stationary state. These are determined 
from averages over the surviving representatives of $10^{3}$-$10^{4}$ 
independent realizations of the SIRS process with the same parameter $(\lambda,\gamma)$, 
beginning with random initial distribution of the $I$'s. 
Surviving sample averages converge to stationary values as $N \rightarrow \infty$. 
To obtain the steady states and critical curve 
we extrapolated quasi-stationary values of finite systems to 
those of the infinite system.

The finite size scaling theory~\cite{marro:1999} can be used to 
obtain the critical curve $\lambda^{z}_{c}(\gamma)$. We can 
assume a scaling function of the surviving probability: 
$P_{t}(I)\sim t^{-\beta/\nu_{\|}}
f((\lambda-\lambda_{c})t^{1/\nu_{\|}})$. 
At criticality, $\lambda=\lambda_{c}(\gamma)$, 
the survival probability of the infection, starting from a single infected 
site, has a power law behavior in time. In the subcritical region, 
it decays exponentially while in the supercritical region it 
reaches non-zero steady state in a short time. The power law behavior 
of the survival probability at criticality enables one to extract 
the critical curve $\lambda^{z}_{c}(\gamma)$ from the time evolution data 
of the SIRS process. This dynamical Monte Carlo simulation 
is reliable when the system size is sufficiently large so that 
the evolution of the system is approximately confined, for the duration of the 
simulation to a region smaller than the size of the system~\cite{grassberger:1979}. 
However we found that this surviving probability oscillates 
wildly when $\gamma$ is small. Because of this 
the dynamical Monte Carlo method is not used to determine the critical curve near $\gamma=0$.

{99}

\newpage

\newpage

\newpage

\newpage

\newpage

\newpage

\newpage

\end{document}